\documentclass[a4paper,twocolumn,altadd,nofootinbib,superscriptaddress,aps,citeautoscript,14pt,
eqsecnum,notitlepage,showkeys]{revtex4-1}
\usepackage{multirow}
\usepackage{graphicx}
\usepackage{bm}
\usepackage{dcolumn}
\usepackage[hidelinks]{hyperref}
\usepackage{gensymb}
\usepackage{amsmath}
\usepackage{dcolumn}    
\usepackage{ifpdf}
\usepackage{amssymb,lineno,amsfonts}
\usepackage{graphicx}   
\usepackage{bm}         
\usepackage{bbm}
\usepackage{mathrsfs}
\usepackage{upgreek}
\usepackage{mathtools}
\usepackage{epstopdf}
\usepackage{setspace}
\usepackage{hyperref}
\usepackage{natbib}
\usepackage{esvect}
\usepackage[usenames,dvipsnames]{xcolor}
\definecolor{med-blue}{RGB}{25,25,112}
\hypersetup{colorlinks, linkcolor={blue},citecolor={blue}, urlcolor={blue}}
\usepackage{times}
\usepackage [english]{babel}
\usepackage [autostyle, english = american]{csquotes}
\MakeOuterQuote{"}

\begin{document}
\title{Weak Ferromagnetism and Time-Stable Remanence in Hematite:  Effect of Shape, Size and Morphology}

\author{Namrata Pattanayak}
\affiliation{Department of Physics, Indian Institute of Science Education and Research, Dr. Homi Bhabha Road, Pune 411008, India}

\author{Arpan Bhattacharyya}
\affiliation{Saha Institute of Nuclear Physics, 1/AF Bidhannagar, Kolkata, India }

\author{Shruti Chakravarty}
\affiliation{Department of Physics, Indian Institute of Science Education and Research, Dr. Homi Bhabha Road, Pune 411008, India}

\author{Ashna Bajpai}
\email[]{ashna@iiserpune.ac.in}
\affiliation{Department of Physics, Indian Institute of Science Education and Research, Dr. Homi Bhabha Road, Pune 411008, India}
\affiliation{Center for Energy Science, Indian Institute of Science Education and Research, Dr. Homi Bhabha Road, Pune 411008, India}

\date{\today}

\begin{abstract}

We have recently established that a number of Dzyaloshinskii-Moriya interaction driven canted antiferromagnets or weak ferrromagnets (WFM) including hematite exhibit an ultra-slow magnetization relaxation phenomenon, leading to the observation of a \textit{time}-\textit{stable} remanence (Phys. Rev. B 96, 104422 (2017)). In this work, our endeavor is to optimize the magnitude of this \textit{time}-\textit{stable} remanence  for the hematite crystallites, as a function of shape size and morphology.  A substantial enhancement in the magnitude of this unique  remanence is observed in porous hematite, consisting of ultra-small nano particles,  as compared to crystallites grown in regular morphology, such as cuboids or hexagonal plates.  This \textit{time}-\textit{stable} remanence exhibits a peak-like pattern with magnetic field, which is significantly sharper in porous sample. The extent and the magnitude of the spin canting associated with the WFM phase can be best gauged by the presence of this  remanence and its unusual magnetic field dependence. Temperature variation of lattice parameters bring out correlations between strain effects that alter the bond length and bond angle associated with primary super exchange paths, which in-turn systematically  alter the magnitude of the \textit{time}-\textit{stable}  remanence.  This study provides insights regarding a long standing problems of anomalies in the magnitude of magnetization on repeated cooling in case of hematite. Our data caps on these anomalies, which we argue, arise due to spontaneous spin canting associated with WFM phase.  Our results also elucidate on why thermal cycling protocols  during bulk magnetization measurements are even more crucial for hematite which exhibits both canted as well as pure antiferromgnetic phase.   

\end{abstract}

\pacs{Valid PACS appear here}

   \begin{figure*}[!t]
\includegraphics[width=1\textwidth]{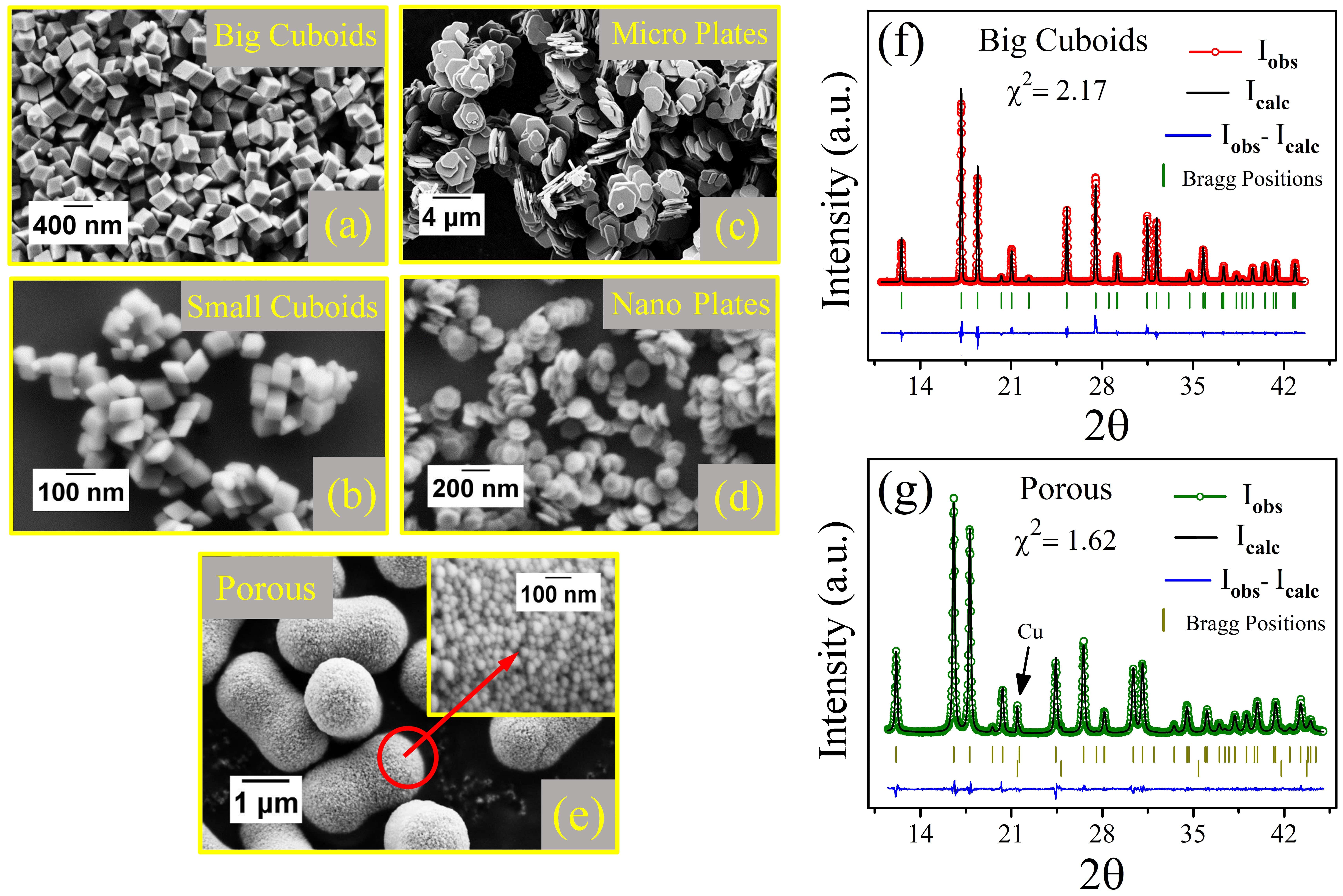}
\caption {\textbf{(a)} SEM images of Hematite (a) big cuboids, (b) small cuboids, (c) micro plates, (d) nano plates and (e) porous sample. (f) - (j) display synchrotron  XRD data for big cuboids and porous Hematite samples respectively, along with Rietveld profile refinement.}
\label{Figure1}
\end{figure*}

\maketitle
\section{Introduction}

Hematite ($\alpha$- Fe$_2$O$_3$) with \textit{T}$_\textit{N}$ $\sim$ 960 K is basically an antiferromagnetic compound, which exhibits a spin reorientation transition at $\sim$ 260 K, known as the Morin transition temperature (\textit{T}$_\textit{M}$) \cite{Morin1}. In a wide temperature range between \textit{T}$_\textit{M}$  and \textit{T}$_\textit{N}$, the spins are still AFM coupled, but exhibit a slight canting phenomenon. This canting, driven by Dzyaloshinskii Moriya Interactions (DMI), leads to a net ferromagnetic moment in the otherwise AFM lattice. The phenomenon, also referred to as “weak ferromagnetism” has been observed not only in hematite but also in a number of isostructural AFMs \cite{Dzy1,Moriya1}. DMI driven spin canting is now of paramount importance due to its fundamental as well as application related aspects in upcoming areas of AFM spintronics and chiral magnets \cite{Onose,Hasan,Binz,Baltz,Jairo,Gomonay,Gayles,Gross}. The onset of WFM in $\alpha$- Fe$_2$O$_3$ is also concurrent with another functionality, namely piezomagnetism, a relatively less explored phenomenon. However PzM was theoretically predicted and experimentally observed in number of such canted AFMs \cite{Dzy2,Romanov1,Romanov2,Romanov3,Romanov4}. PzM relates to the possibility of stress induced moments and therefore holds promising technological implications \cite{Paul,Philip1, Philip2, Kleemann, Binek1}.

   There have been numerous studies on the variation of \textit{T}$_\textit{M}$ as observed in magnetization vs Temperature for hematite in different size and morphology \cite{Dunlop,Amin,Zysler,Mitra}. However, in this work our focus is on exploring the effect of morphology on the remanent magnetization or remanence in hematite. We have earlier observed a unique remanence in not only a number of  symmetry allowed WFM/PzM, including a single crystal hematite \cite{Pattanayak},  but also in some isostructural systems wherein this effect arises due to size and interface \cite{Ashna1, Ashna2}. We have also established that the remanance in these canted AFM has some unique footprints \cite{Pattanayak}, which sets it apart from any other conventional or complex magnetic systems \cite{Morup,Ben1,Ben2,Binder, Mat, Suzuki}. Here the key observation has been an ultra-slow magnetization relaxation phenomenon, resulting in the observation of a time-stable remanence, hereafter reffered to as $\mu$. Hematite is also a unique compound  as it exhibits both AFM and WFM phase as a function of temperature \cite{Dzy1}. Interestingly,  the \textit{time-stable} remanence appears in the WFM phase but it is negligibly small in its pure AFM state, as is observed in single crystal of hematite \cite{Pattanayak}.  
   
   We also find that the magnitude of the \textit{time}-\textit{stable} remanence is small in hematite, as compared to  MnCO$_3$, which has relatively smaller \textit{T}$_\textit{N}$  $\sim$ 30 K. It is known to be a stronger WFM as compared to hematite, for which the \textit{T}$_\textit{N}$ $\sim$ 960 K. Thus comparing the magnitude of $\mu$ in different systems provides insights about the magnitude of spin canting, – a non trivial parameter to estimate. Nano scaling significantly tunes the magnitude of this peculiar remanence \cite{Ashna1,Ashna2,Pattanayak}.   Considering that the spin canting and associated WFM and PzM effects exist in hematite near the room temperature, it is more suitable for practical applications. Therefore it is important to explore shape and size effects to optimize $\mu$ in hematite,  .   
   
   In the present work we report magnetization and remanence measurements in five different samples of hematite using SQUID magnetometry. This includes nano cubes, hexagonal plates and a porous sample. The paper has been organized as follows: The magnetization data on cubic and hexagonal crystallites have been discussed in separate sections to bring forward key size effects, within the same morphology. This is further compared with similar data on a porous sample, consisting of ultra small nano particles.  The detailed structural analysis presented in a separate section brings out correlations between strain effects, lowering of \textit{T}$_\textit{M}$ and its implications on the magnitude of the time-stable remanence.
    
   \section{Experimental Techniques}
The hematite samples have been synthesized by hydrothermal route \cite{Bigcuboids, Nanoplates} (Sup-Info: Text S1). The morphology and size of as synthesized samples are recorded using a Zeiss Ultra plus FESEM, Figs. 1(a)-1(e). The phase purity and crystallinity of the samples have been characterized by using a Bruker D8 advance powder X- ray diffractometer (XRD) with Cu $K_\alpha$ radiation ($\lambda$ = 1.54056 \AA). Temperature variation of synchrotron XRD in the range 300 K - 20 K have been carried out in the BL-18 beamline, Photon Factory, Japan. The diffraction patterns are extensively characterized by Rietveld profile refinement                
 \cite{Young} using FULLPROF software. XRD pattern for two representative samples is shown in Figs. 1(f)-1(g). The magnetization measurements have been conducted using a SQUID magnetometer from Quantum Design. The samples are in the powder form with morphology shown in Figs. 1(a)-1(e).

\subsection*{Experimental Protocol for Magnetization Measurements} 

Magnetization (\textit{M})  as a function of temperature (\textit{T}) is usually measured in Field Cooled (FC) or Zero Field Cooled (ZFC) protocols. In the FC protocol, the sample is typically cooled from above its magnetic transition temperature. In case of hematite, the Neel temperature  \textit{T}$_\textit{N}$ $\sim$ 960 K, however the SQUID data is typically recorded via cooling the sample from 300 K, which is above the Morin transition \textit{T}$_\textit{M}$ in case of hematite.  A representative \textit{M} vs \textit{T} data is shown in Fig. 2(a) for the micro plates.  As mentioned before, the Morin transition \textit{T}$_\textit{M}$, involves a first order spin reorientation transition and marks the onset of WFM state in hematite. For the sake of clarity, the spin configuration in both AFM and WFM state  is schematically shown in Fig. 2(b). The four spin configuration shown in the middle identifies the AFM unit cell with red star mark as inversion center. The spins point along the \textit{c} axis in pure AFM state (below \textit{T}$_\textit{M}$), as schematically shown in the left panel of Fig. 2(b). Above \textit{T}$_\textit{M}$, spins run to basal plane and exhibit spin canting (not shown explicitly) as displayed in the right panel of Fig. 2(b). Considering that the spin reorientation transition at \textit{T}$_\textit{M}$ is known to exhibit thermal hysteresis on heating and cooling cycles, in this work we restrict ourselves to only FC cycles for all the \textit{M} vs \textit{T} data. In the following we discuss the magnetization and corresponding remanence for each type of sample. 

\begin{figure}[!t]
\includegraphics[width=0.49\textwidth]{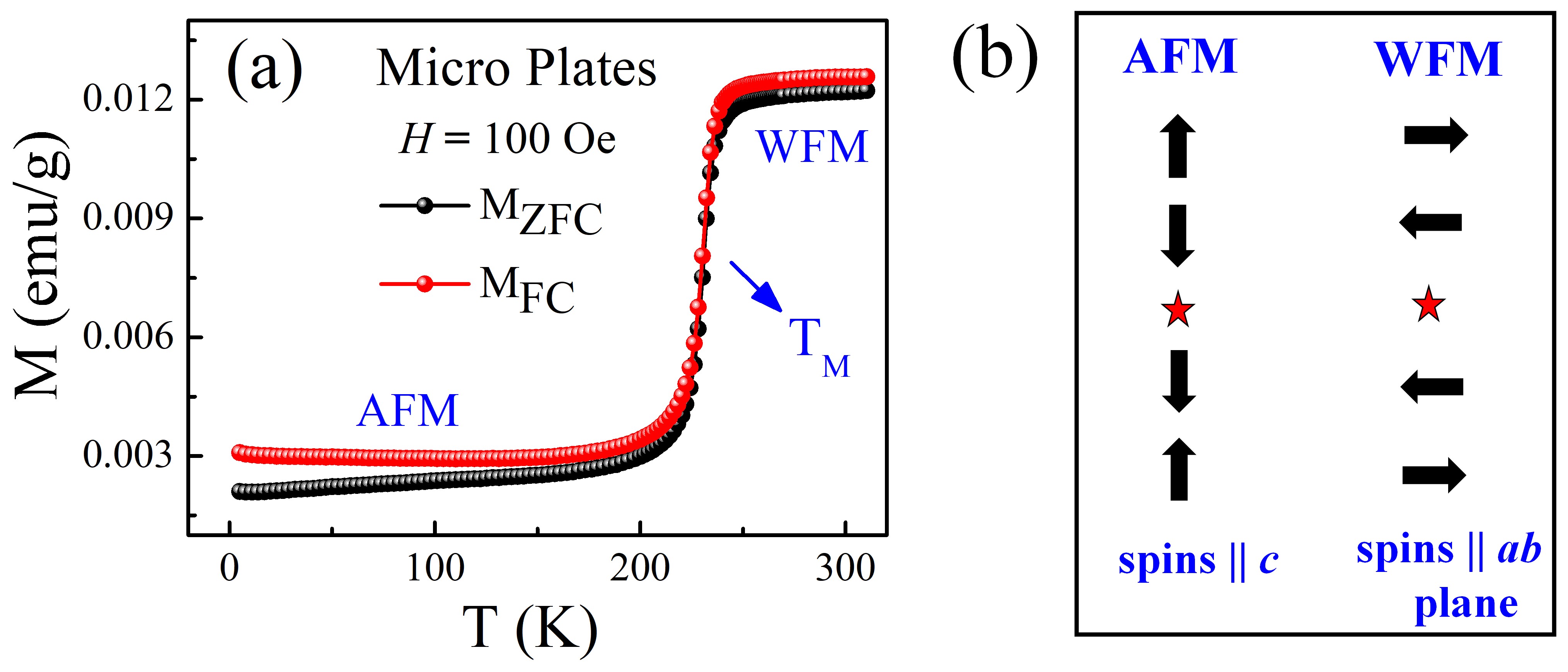}
\caption {\textbf{(a)}  Magnetization as a function of temperature, recorded in FC (red dots) and ZFC (black dots) cycles. The Morin transition, \textit{T}$_\textit{M}$ demarcates the AFM and WFM region intrinsic to Hematite. \textbf{(b)} Depicts the schematic diagram of the typical spin configuration in the AFM and WFM regions in the hexagonal setting with red star as inversion center.}
\label{Figure2}
\end{figure}

\section{Results and Discussions}

\textcolor[rgb]{0,0,1}{\subsection{ Magnetization and Remanence for Porous Hematite: }}

We first discuss the  magnetization and the remanence data in porous sample shown in Fig. 1(e). This sample also serves as a representative for stating the experimental protocol of recording the remanence for all the samples discussed here. Black dots in Fig. 3(a) displays the \textit{M}$_{FC}$  vs \textit{T} data recorded  during cooling the sample from 300 K down to 5 K in presence of \textit{H} = 1 kOe. The \textit{H} is switched off after reaching 5 K for the measurement of $\mu$.  The green dots in Fig. 3(a) shows $\mu$ vs \textit{T}, while warming the sample upto 300 K , while \textit{H}= 0. All the $\mu$ vs \textit{T} data reported in this work has been obtained following the FC protocol.

   Considering the data presented in Fig. 3(a), we observe that  prior to switching off \textit{H}, the magnetization value \textit{M} $\sim$ 0.056 emu/g at \textit{H} = 1 kOe.  After switching off \textit{H}, the magnetization decays to about 50\% of its in-field value ($\mu$ $\sim$ 0.025 emu/g).  As long as the temperature is held constant at 5 K, this remanence exhibits practically no further decay in time.  On increasing the temperature from 5 K to 300 K, we find that the functional form of $\mu$ vs \textit{T} is qualitatively similar to \textit{M} vs \textit{T } for the porous sample, as evident from Fig. 3(a).  The slight hump at 125 K is also indicative of the Morin transition \textit{T}$_M$, which is quite subtle in case of porous samples, as compared to the remanence data obtained on cuboids and hex plates discussed in the latter part of the text.  
	
	\begin{figure}[!t]
\includegraphics[width=0.48\textwidth]{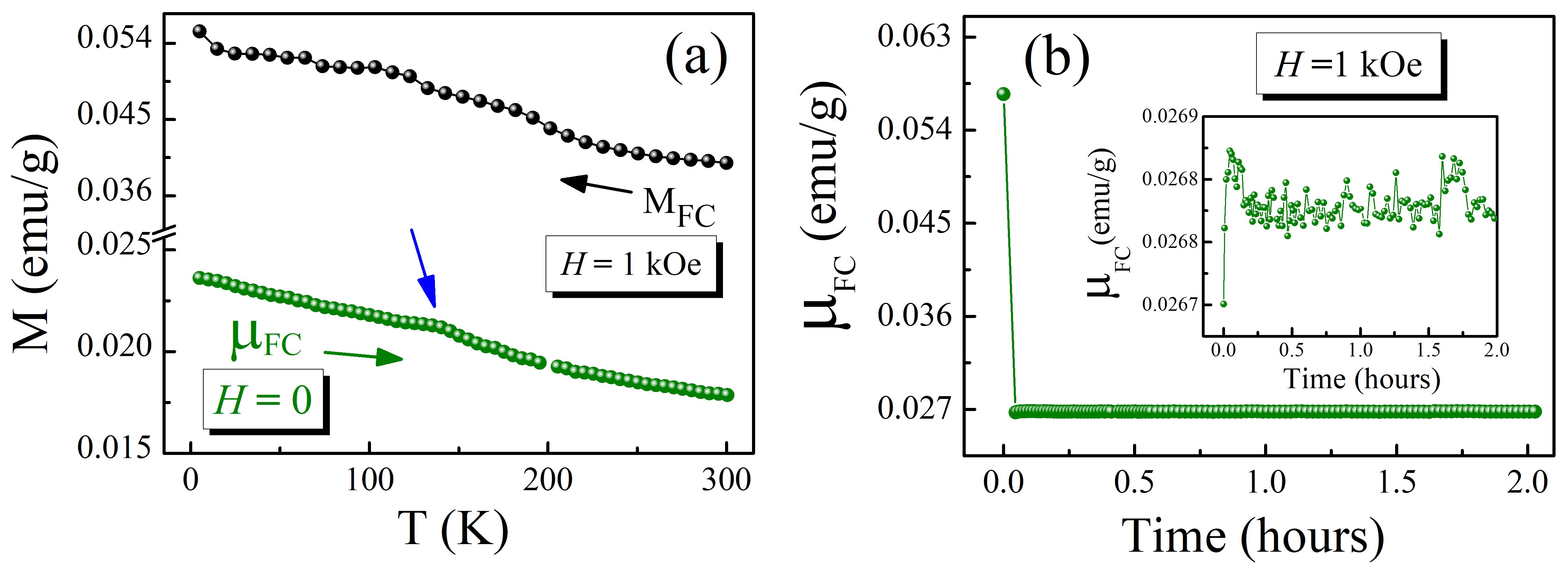}
\caption {\textbf{(a)} \textit{M}$_{FC}$ vs \textit{T} (black dots) measured at \textit{H} = 1 kOe and the corresponding $\mu_{FC}$ vs \textit{T} (green dots) data, measured after removal of \textit{H} = 1 kOe for the porous sample. \textbf{(b)} shows $\mu_{FC}$ as a function of time.The inset in \textbf{(b)} clearly shows that the remanence is almost constant over a time period of 2 hours and hence is \textit{time-stable} in nature.}
\label{Figure3}
\end{figure}
	
We have also measured remanence as a function of time (Fig. 3(b)) to show  that only a part of the remanence is \textit{time}-\textit{stable} in character.  Here again, the remanent state is prepared by cooling the sample in \textit{H} = 1 kOe. A single  data point  at 5 K for  in-field magnetization is shown in  the main panel of Fig. 3(b). On switching off the \textit{H}, the 50\% of in-field magnetization  decays instantaneously,  as is evident from the sudden drop in Fig. 3(b).  However a part of the remanence exhibit almost no decay with time. Its magnitude changes by less than 0.1\% over the time period of two hours, as is also evident from the inset of Fig. 3(b). This part of the remanence, which is the subject matter of investigation here, is fairly constant in time and can be termed as quasi static or \textit{time}-\textit{stable}.

\textcolor[rgb]{0,0,1}{\subsection{ Magnetization and Remanence for Cuboids of Hematite:}}

In this section we present the \textit{M} and corresponding $\mu$ data for two different cuboids of hematite, with side-lengths of 60 and 200 nm respectively. The SEM for these two samples has been displayed in Figs. 1(a) and 1(b). \textit{M} vs \textit{T} at 100 Oe and 1 kOe is  compared  for both these samples in Fig. 4(a) and 4(b). Similar data recorded for \textit{H} = 10 and 50 kOe has been given as Sup-Info: Fig. S1.  As inferred from \textit{M} vs \textit{T} data, the \textit{T}$_\textit{M}$ is seen to reduce for small cuboids (red dots)  in comparison with big cuboids ( black dots). This feature is consistent with the previous reports \cite{Dunlop,Amin,Zysler,Mitra}.  We also note that for \textit{H} = 100 Oe corresponding to small cuboids, magnetization is significantly larger in WFM region. However for \textit{H} = 1 kOe, the magnitude of magnetization is similar in both regions, below and above \textit{T}$_\textit{M}$.  

	\begin{figure*}[!t]
\includegraphics[width=1\textwidth]{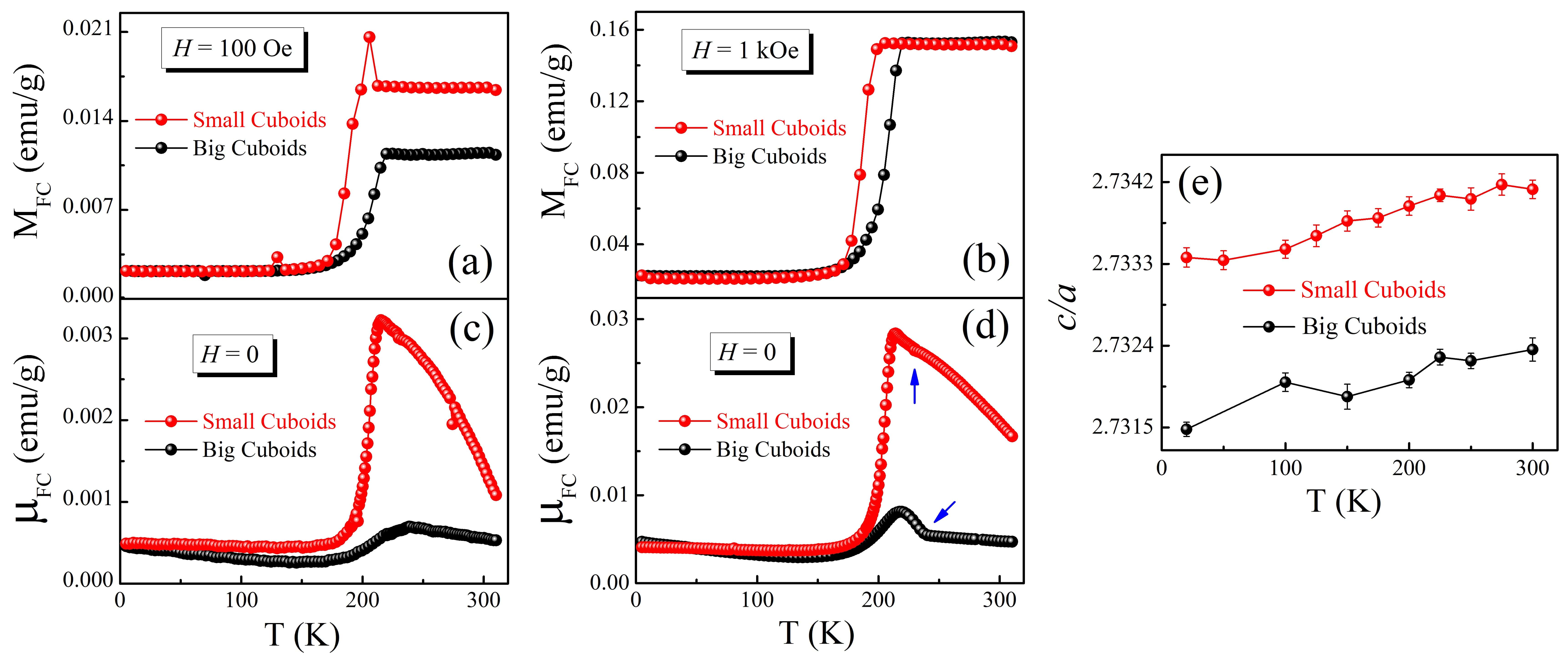}
\caption {\textit{M}$_{FC}$ vs \textit{T} for  big cuboids (black dots) and small cuboids (red dots) obtained at different magnetic fields of \textbf{(a)} \textit{H} = 100 Oe, \textbf{(b)} \textit{H} = 1 kOe. \textbf{(c)} and \textbf{(d)} compares the corresponding $\upmu_{FC}$ vs \textit{T} data. \textbf{(e)} Compares the temperature variation of  c/a ratio of both samples. Here the lattice parameters \textit{c} and \textit{a} are determined from the Rietveld profile refinement of the synchrotron XRD data.}
\label{Figure4}
\end{figure*}

The corresponding $\mu$ vs \textit{T} recorded in warming cycle for the \textit{M} vs \textit{T} runs is shown in Figs. 4(c) and 4(d) respectively.  The $\mu$ vs \textit{T} data clearly marks the onset of \textit{T}$_M$, similar to what is observed in routine \textit{M} vs \textit{T}. However, the magnitude of $\mu$  is higher for small cuboids above \textit{T}$_\textit{M}$,  for \textit{H} = 100 Oe as well as for \textit{H} = 1 kOe. The magnitude of $\mu$ is similar below \textit{T}$_\textit{M}$for the remanent state prepared either at \textit{H} = 100 Oe or \textit{H} = 1 kOe. Another striking difference is the temperature dependence of remanence in the WFM region. It is evident that for cuboids, the remanence decays at much faster rate with temperature above the \textit{T}$_\textit{M}$. Below \textit{T}$_\textit{M}$, the functional form of $\mu$ vs \textit{T} is more or less similar, irrespective of the size of cuboids. The  $\mu$ vs \textit{T} data in both the samples  in WFM region exhibit some subtle anomalies,  indicating signatures of double transition (marked by arrows). These features in $\mu$ vs \textit{T} are not as prominent in the in-field  \textit{M} vs \textit{T} data, highlighting the importance of remanent magnetization to uncover the subtleties of associated with WFM phase of hematite.  This feature appears to be related to DMI associated with two possible symmetry allowed \textbf{D}(\textbf{S}$_i$ X \textbf{S}$_j$) type of interactions between AFM planes in hematite. 

Fig. 4(e) shows the \textit{c}/\textit{a} ratio, where \textit{c} and \textit{a} are the lattice parameters of hematite, as derived from Rietveld profile refinement of the synchtrotron XRD data. The strain effects in lattice parameters are larger for small nano cubes, consistent with relatively larger magnitude of remanance upon nano scaling, while keeping the morphology same. It is evident that the magnitude of remanance is larger when \textit{T}$_\textit{M}$ is reduced and strain effects are larger, such as the case of small cuboids. 

\textcolor[rgb]{0,0,1}{\subsection{ Magnetization and Remanence for Hexagonal Plates of Hematite:}}

Fig. 5(a)-5(c) shows \textit{M} vs \textit{T} for the hematite samples consisting of nano plates (pink dots) and micro-plates (blue dots), with the morphology of individual plates being hexagonal in both cases.  The side length (and thickness)  of individual nano and micro plates is 70 nm (15 nm) and 1.5 $\upmu$m (300 nm) respectively, as shown in Figs. 1(c) and 1(d) respectively.  The magnetization and the corresponding remanence at 100 Oe and 1 kOe is  compared  for both these samples in Fig. 5(a) and 5(b). Similar data recorded for \textit{H} = 10 and 50 kOe has been given as Sup-Info: Fig. S2. As expected the \textit{T}$_\textit{M}$ shifts towards lower temperature for nano plates. The magnitude of remanance is again higher for nano plates for \textit{H} = 1 kOe, even though the corresponding \textit{M} values are similar for both the samples. Here the remanance falls at a relatively  faster rate with temperature on the either side of \textit{T}$_\textit{M}$. For hex plate morphology, the anomaly in lattice parameters above the Morin transition is more prominent than what is seen in cuboids (Fig. 5(d)). The data suggests that upon nano scaling, due to large surface to volume ratio, the canting angle may be larger but the effects is relatively less robust as a function of temperature.

\begin{figure*}[!t]
\includegraphics[width=1\textwidth]{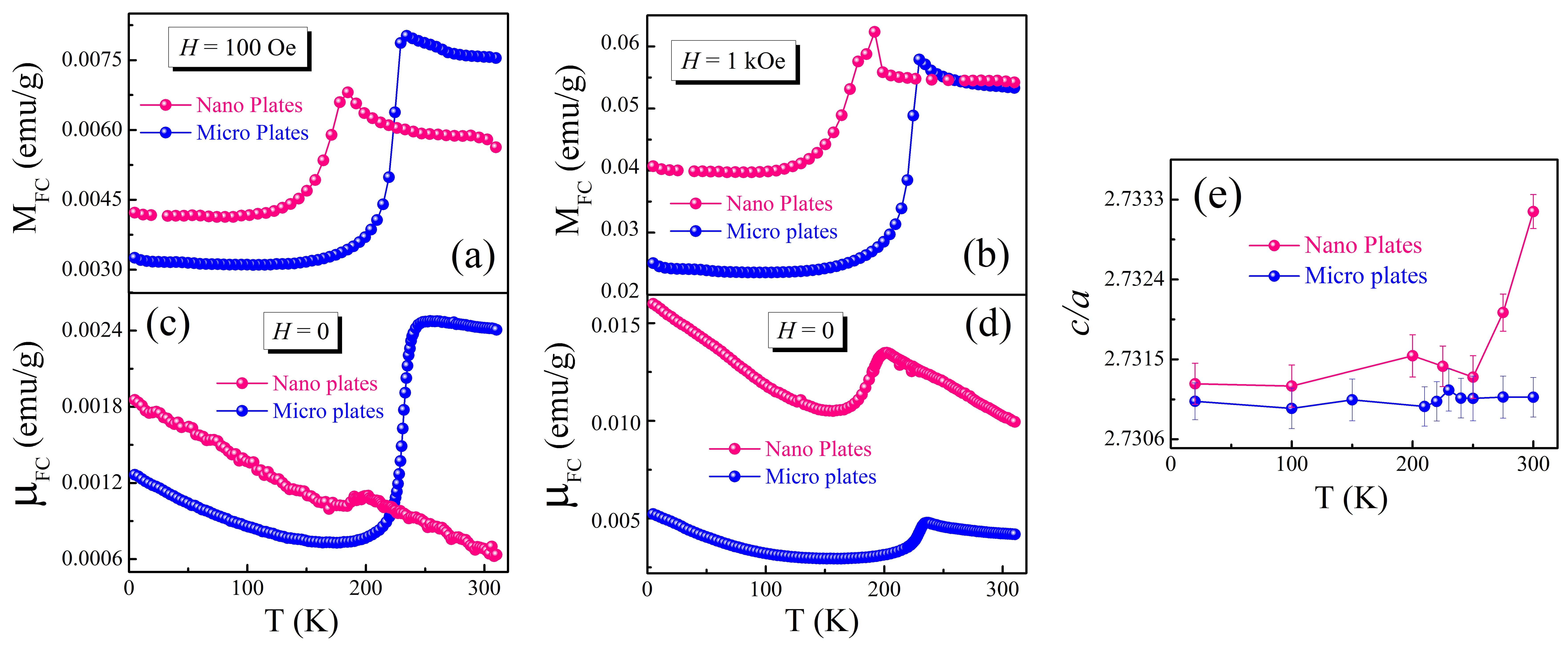}
\caption {\textit{M}$_{FC}$ vs \textit{T} for  micro plates (blue dots) and nano plates (pink dots) obtained at different magnetic fields of \textbf{(a)} \textit{H} = 100 Oe, \textbf{(b)} \textit{H} = 1 kOe. \textbf{(c)} and \textbf{(d)} compares the corresponding $\mu_{FC}$ vs \textit{T} data. \textbf{(e)} Compares the temperature variation of  \textit{c}/\textit{a} ratio of both samples, determined from the Rietveld profile refinement of the synchrotron XRD data.} 
\label{Figure5}
\end{figure*} 

\textcolor[rgb]{0,0,1}{\subsection{ Remanence as a function of (cooling ) H: cuboids, plates and the porous sample}}

After highlighting a few key observation regarding the nature of remanence  upon down scaling,  while keeping the morphology same, we now compare the magnetic field dependence of the remanence for all five samples.  Focusing on the \textit{H} dependence of remanence, we consider its magnitude at 5 K and also at 300 K. These values are chosen up from  various $\mu$ vs \textit{T} runs, in which the remanent state is prepared in different (cooling) magnetic field for each sample. As shown in Figs. 6(a) and 6(b), the magnitude of $\mu$ peaks when it is prepared at the \textit{H} = 1 or 10 kOe.  However, comparing data on all five  morphologies, this peak value is significantly higher for the porous sample at 5 K (as well as at 300 K) . 

\begin{figure*}[!t]
\includegraphics[width=1\textwidth]{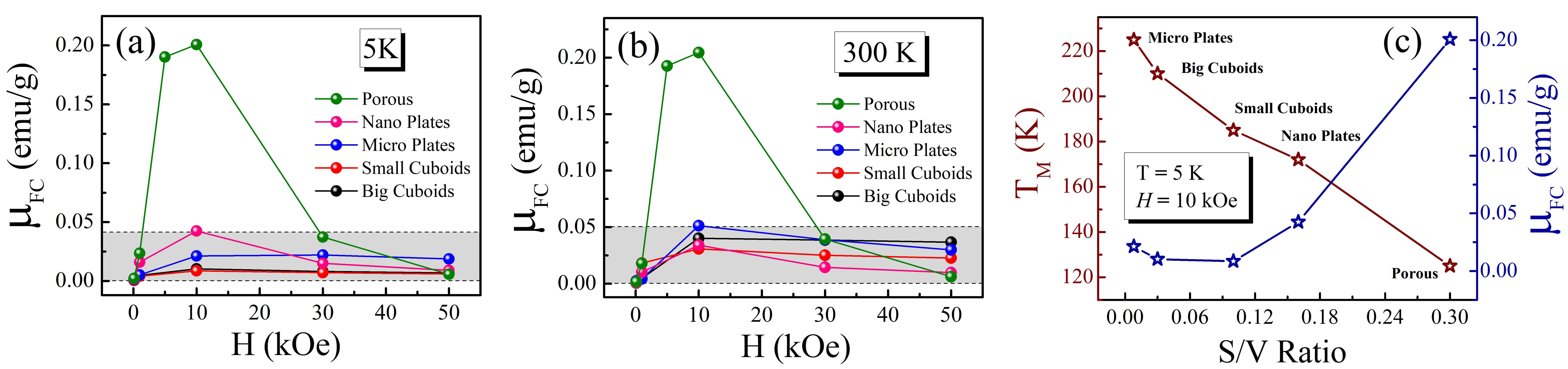}
\caption {The magnetic field dependence of remanence  for all samples  \textbf{(a)} 5 K and \textbf{(b)} 300 K, depicting that the peak value of $\mu$ is obtained for \textit{H} = 1-10 kOe. This also highlights the magnitude of $\mu$ is significantly larger for the porous sample. \textbf{(c)} The variation of the Morin transition temperature (left axis) as well as the peak value of $\mu$ (right axis) as a function of the S/V ratio of the hematite samples.}
\label{Figure6}
\end{figure*}

As highlighted by a grey area in both the figures, the magnitude of $\mu$ for cuboids and hex plates, irrespective of their size, is roughly in the range of 0.05 emu/g, whereas it is atleast 4 times (0.21 emu/g) for the porous sample. Apart from this significant enhancement in $\mu$ for the porous sample, the peak like feature in $\mu$ vs \textit{H} is also much sharper, as compared to the hematite samples with regular morphology of cuboids / hex plates. This peculiar magnetic field dependence of  $\mu$ also sets it apart from other conventional or complex magnets \cite{Morup,Ben1,Ben2,Binder, Mat, Suzuki}.  It is also clear that the magnitude of $\mu$ is significantly tuned by particle’s surface to volume ratio.

The \textit{T}$_\textit{M}$ and the magnitude of $\mu$ as a function surface to volume ratio  is shown in Fig. 6(c), covering all five samples. Consistent with previous reports, nano scaling leads to systematic reduction in the \textit{T}$_\textit{M}$ as is shown in left axis of Fig. 6(c). Though not shown here, this is associated with reduction of  corresponding \textit{T}$_\textit{N}$.  Reduction in \textit{T}$_\textit{N}$ implies weakening of  basic AFM interactions driven by super exchange. This should lead to a larger spin canting effects and hence the associated net FM moment in otherwise AFM lattice. The magnitude of \textit{time-stable} remanence reflects this feature in a clear fashion as is evident from the right axis of Fig. 6(c). This is also consistent with previous report, the magnitude of remanance increases with decrease in \textit{T}$_\textit{M}$.

\begin{figure*}[!t]
\includegraphics[width=1\textwidth]{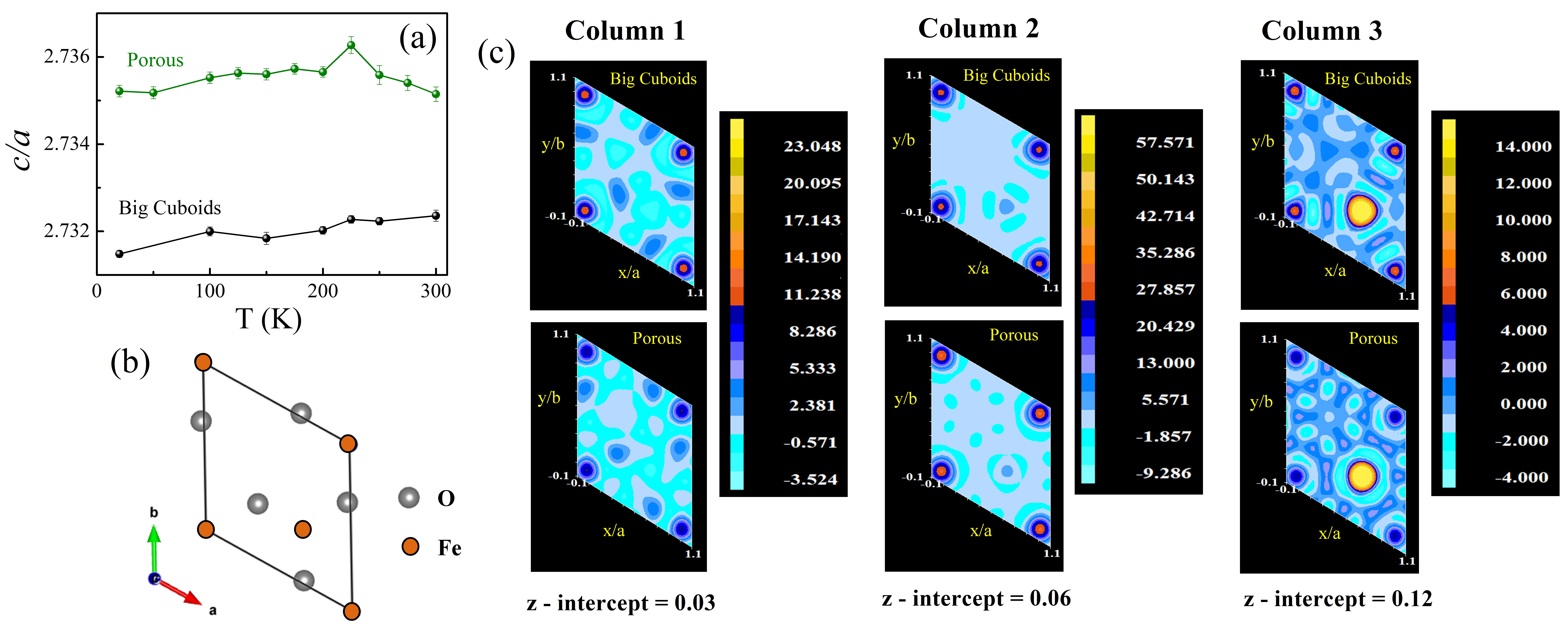}
\caption {\textbf{(a)} Compares the temperature variation of the \textit{c}/\textit{a} ratio, big cuboids (black dots) and porous sample (green dots). \textbf{(b)} Depicts a slice of hexagonal unit cell of hematite (top view). Here the xy plane, generated using the vesta programme displays the oxygen atom (grey balls) and the Fe atom (green balls). \textbf{(c)} Shows the electron density (ED) maps of  big cuboids and porous sample obtained by the Fourier transformation of synchrotron XRD using Rietveld refinement. The comparative study of ED maps depict the strain effects arising due to nano scaling.} 
\label{Figure7}
\end{figure*}

    It is clear from Figs. 6(a) and 6(b), that following a FC protocol in a routine \textit{M} vs \textit{T} measurement, once a sample is cooled in a certain \textit{H} there exists a \textit{time}-\textit{stable} $\mu$, the magnitude of which is related to the \textit{H} used in previous run.  The magnitude of $\mu$ is small at very low \textit{H} as well as at very high \textit{H}.  There is an optimum \textit{H} for which the magnitude of \textit{time}-\textit{stable} $\mu$ is maximum for each sample, leading to a peak like pattern in $\mu$ vs \textit{H}, as is evident from Figs. 6(a) and 6(b). The presence of this \textit{time-stable} $\mu$ as well as its \textit{H} dependence explains the discrepancies on magnetization data on repeated cooling\cite{Romanov1,Sendonis}. Depending on the magnitude of the \textit{H} used while cooling and on the morphology of the sample, this contribution can be 50- 90\% of the in-field \textit{M} value \cite{Pattanayak,Kapoor}. This contribution to magnetization comes from the spontaneously canted AFM domains and it is best gauged by the presence of \textit{time}-\textit{stable} $\mu$. In case of hematite, cooling from above the \textit{T}$_\textit{M}$ in presence of \textit{H}, the total magnetization is driven by spontaneously canted domains in the direction of \textit{H}. This part is related to the presence of \textit{time-stable} remanence. When \textit{H} is above a critical value, only Zeeman  and other routine  energy terms dominates \cite{Blundell,Joy}, which exhibit instantaneous time decay , such as shown in Fig. 3(b). For large \textit{H}, the magnitude of \textit{time}-\textit{stable} remanence  is small. Thus for each sample, there is an optimum \textit{H}, for which the magnitude of this \textit{time}-\textit{stable}  remanence reaches its maximum value.

    Apart from the optimum \textit{H} for maximizing the magnitude of $\mu$, heating cooling protocol also play an important role. Heating in presence of \textit{H} from below the \textit{T}$_\textit{M}$ should enable larger number of WFM domains to point in the direction of \textit{H}. For sake of consistency, we have only shown the remanent state following FC protocol in this work. Remanent states prepared in both FC and ZFC protocols have been discussed in \cite{Kapoor}. Overall, we infer that the uncertainties on \textit{M} vs \textit{T} data in hematite crucially relate to the WFM phase and can be understood by considering the presence of \textit{time}-\textit{stable} $\mu$. On microscopic level, it should relate to the number and type of canted domains.  Thus heating cooling cycles as well as the magnitude of \textit{H} applied during magnetization measurements profoundly effects the number of WFM domains, which accordingly reflect in the magnitude of  time-stable part of remanence. This also  relates to a unique pinning mechanism \cite{Pattanayak}. 
    
       In case of the porous sample, the peak like effect in $\mu$ vs \textit{H} is sharper (smaller FWHM) as compared to all other morphologies of the hematite. Its \textit{T}$_\textit{M}$ is also significantly smaller as compared to all other morphologies.  This implies the weakening of superexchange paths, that enables larger extent of spin canting. We have earlier observed similar features in MnCO$_3$ which has a relatively smaller \textit{T}$_\textit{N}$ than hematite. In MnCO$_3$,  the FWHM of $\mu$ vs \textit{H} is sharper, the \textit{T}$_\textit{N}$ smaller and the magnitude of remanence higher than bulk hematite\cite{Pattanayak}. This correlations implies that comparing the magnitude of time-stable remanence can provide insights about the nature of spin canting and its extent.

    \textcolor[rgb]{0,0,1}{\subsection{ Strain effects in lattice parameters and time-stable $\mu$: }}
		
		 Focusing on one system, such as the case of hematite, the possibility of nano scaling leads to strain effects that are known to profoundly affect the super exchange paths.  This should effect the magnitude as well as the angle of spin canting. Thus information about lattice parameters, bond angle and bond length due to strain effects and its correlation  with  \textit{time}-\textit{stable} $\mu$ can provide crucial information about the WFM phase.  Figs. 4(e) and 5(e) bring out that strain effects are larger in  small cuboids or thin hex plates , however the differences in the corresponding \textit{T}$_\textit{M}$ (in the range of 170-200 K) in these four samples is not as widely different as compared to the porous sample (\textit{T}$_\textit{M}$ $\sim$ 125 K). The same feature reflects in the magnitude of corresponding time-stable $\mu$ as is evident from Fig. 6(c). From the shaded grey rectangle in Figs. 6(a) and 6(b), it is also clear that cubes and plates have similar magnitude of remanence, as compared to the porous sample. Therefore for the sake of conciseness in the discussion, we choose big nano cubes as a representative of hematite crystallites of regular shape shown Figs. 1(a)-1(d) and compare it with porous hematite, as far as the detailed structural analysis is concerned.

		\begin{figure*}[!t]
\includegraphics[width=1\textwidth]{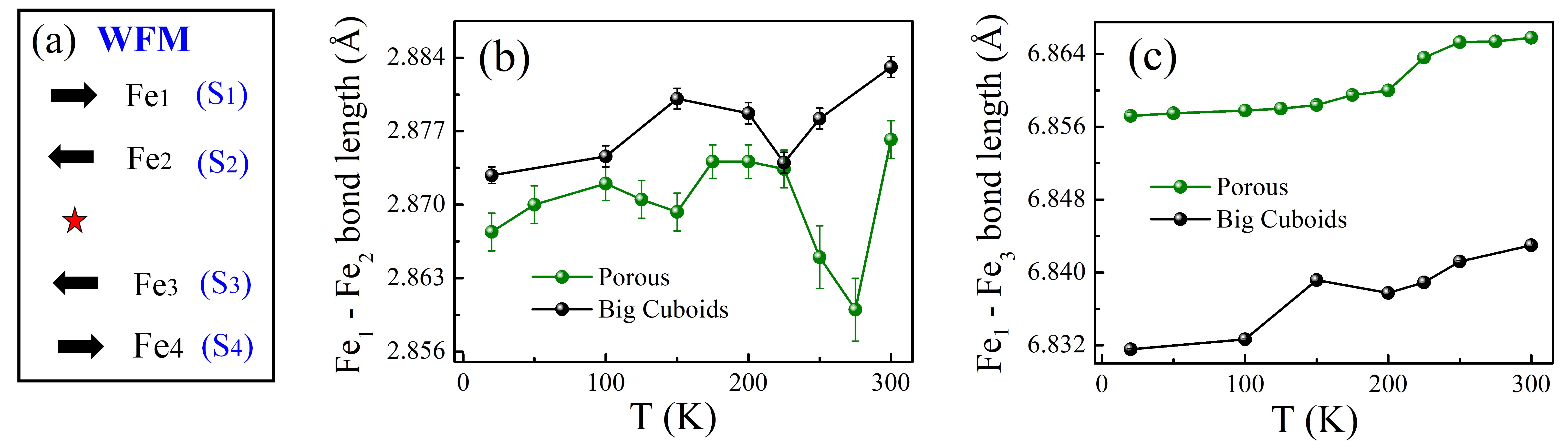}
\caption {The temperature variation of the bond lengths between \textbf{(a)}  S$_1$ and S$_2$ and \textbf{(b)} S$_1$ and S$_3$ along the c-axis of the hexagonal unit cell (as shown in Fig. 2b) in the big cuboids (black dots) and porous (green dots) respectively.}
\label{Figure8new}
\end{figure*} 
	
    In Fig. 7(a), the \textit{c}/\textit{a} lattice parameter of big cuboid is compared with the porous sample. Consistently, the strain effects are significantly larger in the porous sample.  To further investigate this feature, we present electron density (ED) maps of  big cuboids and porous sample obtained from the Rietveld refinement of their respective synchrotron XRD data. A slice of hexagonal unit cell of hematite in the xy plane is shown in Fig. 7(b). Here the orange and the grey balls represent the Fe and oxygen atoms respectively. The four corner Fe atoms lie on the same plane.  The middle Fe atom lies little above the plane containg four corner Fe atoms. There are four such layers of corner Fe atoms (two above and two below the center of inversion) that are present in a unit cell of hematite. Traversing in the z direction, the ED maps obtained at different z intercepts are shown in coulmn 1-3 Fig. 7(c). The scale bar corresponds to the electron density of Fe  and O atoms is shown adjacent to each column. For each coulmn, the top  panel displays ED map for the big cuboid and the bottom panel is for the porous sample.

     As shown in Fig. 7(c), at the z intercept of 0.03 (coulmn -1) the first layer of four corner Fe atoms  just start to appear. Here the ED is larger for the cuboid as compared to the porous sample (lower panel) . In column -2, at the z intercept of 0.06,  the ED around the corner Fe atoms in both the samples approaches its maximum but the ED in porous sample is larger.  At the z intercept of 0.12 the four corner Fe atoms just about to disappear. Similar to the z intercept of 0.03 (column-1) the ED of corner Fe atoms is smaller in Porous sample. It is also apparent for the porous sample  the ED around the Fe atoms is largely confined at the center, whereas the ED of Fe atoms in the big cuboids is more uniformly distributed along the z direction. These data pictorially represent the strain effects, which are the consequence of nano scaling. These strain effects are clearly larger for the porous sample. The strain effects microscopically relate to change in bond length and bond angles that ultimately dictate the variations in \textit{T}$_\textit{N}$ and \textit{T}$_\textit{M}$. The major superexchange paths that lead to basic AFM interaction in the hexagonal unit cell of hematite \cite{hill} and the corresponding bond length and bond angle have been given in Sup-Info (Text S3 and Fig. S3). Here the bond angle and bond length increases for the porous sample (Figs. S3(b) and S3(c)), implying weakening of the primary superexchange paths. This observation is also consistent  with the lowering of    
     \textit{T}$_\textit{M}$ and larger magnitude of \textit{time-stable} $\mu$ in the porous sample.

			In Fig. 8(a) we schematically show spin configuration in WFM state for the hematite unit cell containing four Fe atoms.  These four Fe atoms are located along the 111 direction of rhomohedral unit cell of hematite,  considered by Dzyaloshiskii (which is equivalently the c direction of hexagonal unit cell)  
			\cite{Dzy1,Rollmann}. For the sake of comparison, we restrict ourselves to these 4 Fe atoms, with spin designated as S$_1$ to S$_4$, shown in the bracket in Figure 8a.  We first plot the bond lengths between the  AFM coupled pairs which are symmetry allowed for DMI driven canting \cite{Dzy1,Moriya1,Birss}. Fig. 8(b) shows the variations in the bond length as a function of temperature,  corresponding to the spin pairs S$_1$ and S$_2$. This  exhibits  anomalous features, especially two broad humps for both the samples. Spin Pair S$_1$ and S$_2$ should be equivalent to spin pair S$_3$ and S$_4$ and correspond to the AFM planes below the inversion center\cite{Dzy1,Moriya1}.  The associated DMI driven coupling should be \textbf{D}$_1$σ$_2$(\textbf{S}$_1$ X \textbf{S}$_2$) and \textbf{D}$_3$σ$_4$(\textbf{S}$_3$ X \textbf{S}$_4$) respectively. As required by symmetry considerations, S$_2$ and S$_3$ should point in the same direction for DMI to occur,  as  is shown schematically in Fig. 8(a). Here D$_1$σ$_2$ and D$_3$σ$_4$ should be related such that the canting is consistent with symmetry considerations, giving rise to finite net FM moment associated with DMI driven canting. These adjacent planes should be the primary cause of net FM moment.

     Looking into the anomalous features in the temperature variation of bond length corresponding to spin pair S$_1$ and S$_2$,  we also present the bond length corresponding to spin pair S$_1$ and S$_3$, (which should be equivalent to S$_2$ and S$_4$). This is possibly the source of secondary DMI driven coupling. The corresponding  bond length between spin pair S$_1$ and S$_3$ also shows a change of slope as a function of temperature. Though more careful microscopic measurements are needed to confirm this, but we propose that this secondary DMI driven coupling should also be taken into account.  Especially the temperature dependence of individual AFM coupled spin pairs  needs to be explored more carefully, looking into  the anomalous strain effects in the lattice parameters, as observed here. This may also enable one to understand the double transition such as seen in case of cuboids in Fig. 4(c) and 4(d).

\section{Conclusion}
  In conclusion, we have presented remanent magnetization data in  various hematite samples. The remanence data on cuboids and hexagonal plates are compared, so as to isolate the key size effects while keeping the morphology same. We have further compared remanence data on  these regular shaped crystallites with a porous sample, consisting of ultra thin nano particles of hematite. In all these samples, we observe a part of remanence, which is \textit{time}-\textit{stable} in character and associated with  the Dzyaloshinskii Moriya Interaction driven spin canting.  For each type of hemtite sample, the optimum magnetic field at which the magnitude of this \textit{time}-\textit{stable} remanence maximizes is determined. This remanence exhibits a peak like pattern as a function of magnetic field, which is sharpest in the case of porous hematite. The height and the width of this peak provides insights about the extent of spin canting associated with the WFM phase.  The temperature variation of remanence data and the lattice parameters obtained by synchrotron X ray diffraction data  bring out a clear correlation between the  extent of spin canting, the Morin transition temperature and the magnitude of the \textit{time-stable} remanence, which is found to be significantly larger in porous sample. The electron density map determined from the Rietvel profile refinement of the XRD data also confirm the anomalous strain fields,  which reflect in the bond angle / bond lengths and correlate with the magnitude of the \textit{time}-\textit{stable} remanence. The data also puts an upper limit  to the anomalies related to magnetization of hematite  on repeated cooling. Presence of this \textit{time}-\textit{stable} remanence  and its peculiar magnetic field dependence explains this ambiguity.  We propose that the presence of this \textit{time}-\textit{stable} remanence with its unique magnetic field dependence  provides a means to distinguish canted antiferromagnets from normal ones.
   
  \section{Acknowledgments} 
  Authors thank Sunil Nair (IISER Pune) for SQUID magnetization measurements. AB acknowledges Department of Science and Technology (DST), India for funding support through a Ramanujan Grant and the DST Nanomission Thematic Unit Program. Authors thank the DST and Saha Institute of Nuclear Physics, India for facilitating the experiments at the Indian Beamline, Photon Factory, KEK, Japan.  



\bibliographystyle{apsrev4-1}
\bibliography{Bibliography}

\end{document}